\documentclass[11pt,twoside]{article}
\usepackage{newpasp}
\usepackage{epsf}
\index{Turnshek, D.}
\index{Rao, S.}
\index{Lane, W.}
\index{Monier, E.}
\index{Nestor, D.}
\index{Bergeron, J.}
\index{Smette, A.}

\begin{document}

\title{Damped Lyman-Alpha Galaxies}
\vspace{-3pt}
\author{D. Turnshek, S. Rao}
\vspace{-2pt}
\affil{Dept. of Physics and Astronomy, The University of Pittsburgh}
\vspace{-3pt}
\author{W. Lane}
\vspace{-2pt}
\affil{Kapteyn Astronomical Institute}
\vspace{-3pt}
\author{E. Monier}
\vspace{-2pt}
\affil{Ohio State University}
\vspace{-3pt}
\author{D. Nestor}
\vspace{-2pt}
\affil{Dept. of Physics and Astronomy, The University of Pittsburgh}
\vspace{-3pt}
\author{J. Bergeron}
\vspace{-2pt}
\affil{ESO Garching}
\vspace{-3pt}
\author{A. Smette}
\vspace{-2pt}
\affil{NASA GSFC}

\vspace{-2pt}
\begin{abstract}

Some results from an imaging program to identify low-redshift
($0.09<z<1.63$) damped Ly$\alpha$ (DLA) galaxies are
presented. The standard paradigm that was widely accepted
a decade ago, that DLA galaxies are the progenitors of
luminous disk galaxies, is now being seriously challenged.
The indisputable conclusion from imaging studies at
low redshift is that the morphological types of DLA galaxies
are mixed and that they span a range in luminosities and
surface brightnesses.
\end{abstract}

\vspace{-25pt}
\section{Introduction}
\vspace{-5pt}

Quasar damped Ly$\alpha$ (DLA) absorbers are widely recognized as
important probes of galaxy formation because they contain the bulk
of the cosmological mass of observable neutral gas in the Universe.
Consequently, studies of ``DLA galaxies'' are potentially among the most
important tools for directly studying the conversion of gas into stars
on a cosmological scale.  In particular, quasar absorption-line analysis,
coupled with the results from follow-up galaxy identification studies,
uniquely offers the opportunity to simultaneously investigate the gaseous
and stellar components of cosmologically distant galaxies, something which
optical/IR galaxy surveys by themselves do not directly permit.
The standard paradigm for over a decade has been that DLA systems arise
in luminous HI disk galaxies, but there are now direct
results which contradict this. Observations are more
consistent with the idea that DLA absorption arises in Giant Hydrogen
Clouds associated with various types of galaxies or protogalaxies,
before the complete conversion of gas into stars (Khersonsky \&
Turnshek 1996).

\begin{figure}[t!]
\plotone{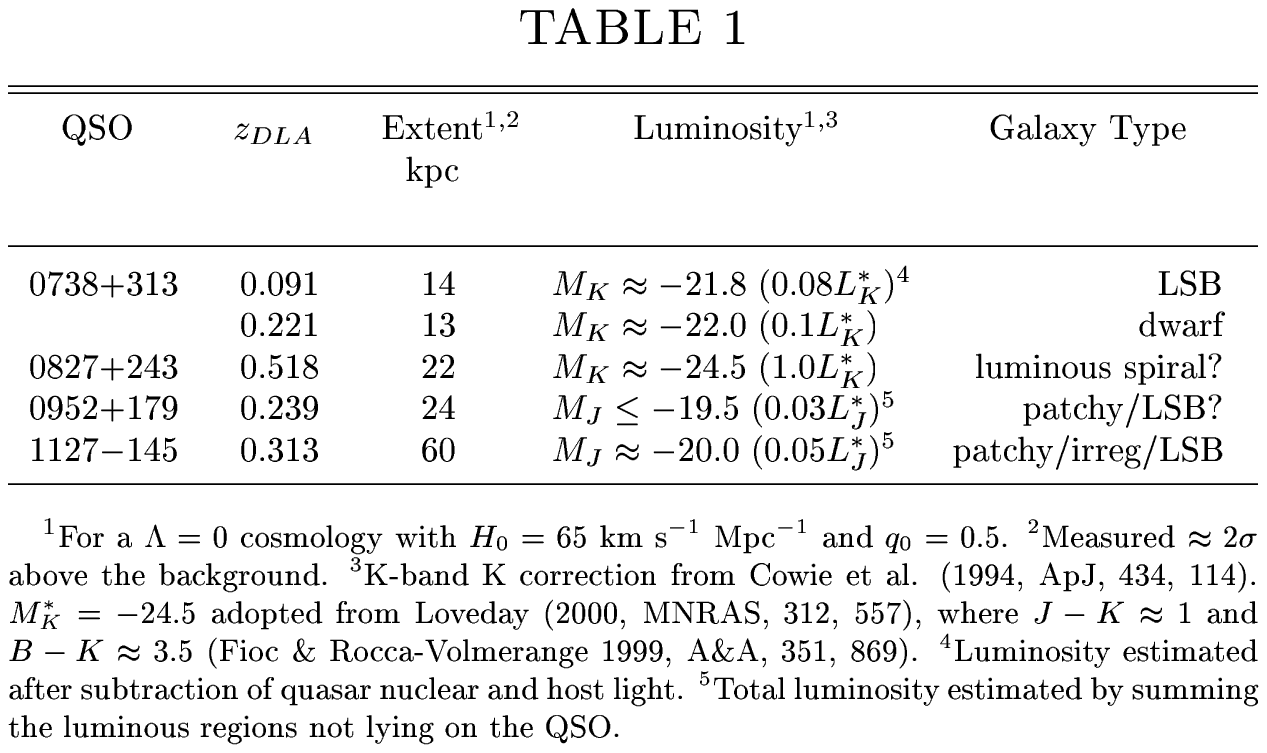}
\vspace{-30pt}
\end{figure}

\vspace{-10pt}
\section{Results}
\vspace{-5pt}

Recent imaging studies with $HST$-WFPC2 have begun to show that the
host galaxies of confirmed low-redshift DLA absorption lines
have a wide variety of
morphological types. In one case, towards 3C336, there is no optical
evidence for the DLA galaxy despite very deep observations (Steidel et
al. 1997). In other cases (Le Brun et al. 1997) the DLA galaxies are
likely to be spirals (towards 3C196 and Q1209+107), amorphous, low
surface brightness galaxies (towards PKS1229$-$021 and 3C286), and
compact objects (towards EX0302$-$223 and PKS0454+039). 
These galaxies have redshifts in the interval $0.395\le z \le 1.010$.

Here we present further evidence for this mixed population from
ground-based infrared observations of five of the lowest redshift DLA
systems discovered in the survey of Rao \& Turnshek (2000;
see also Rao, these proceedings). Table 1
summarizes the properties of these five DLA galaxies and Figure 1
shows the images. Other pieces of evidence which challenge the
luminous-disk hypothesis include: (1) the low-redshift DLA HI column density
distribution does not fall off as $\sim N^{-3}$ at high column
density, as would be expected for disks (Rao \& Turnshek 2000), (2)
the elemental abundances are apparently not evolving toward values
typical of the solar neighborhood (Pettini et al. 1999), and (3)
the kinematic properties of metal lines associated with
high-redshift DLA systems can reasonably be interpreted as fragments
of infalling gas (Haehnelt et al. 1998).

\begin{figure}[t!]
\plotone{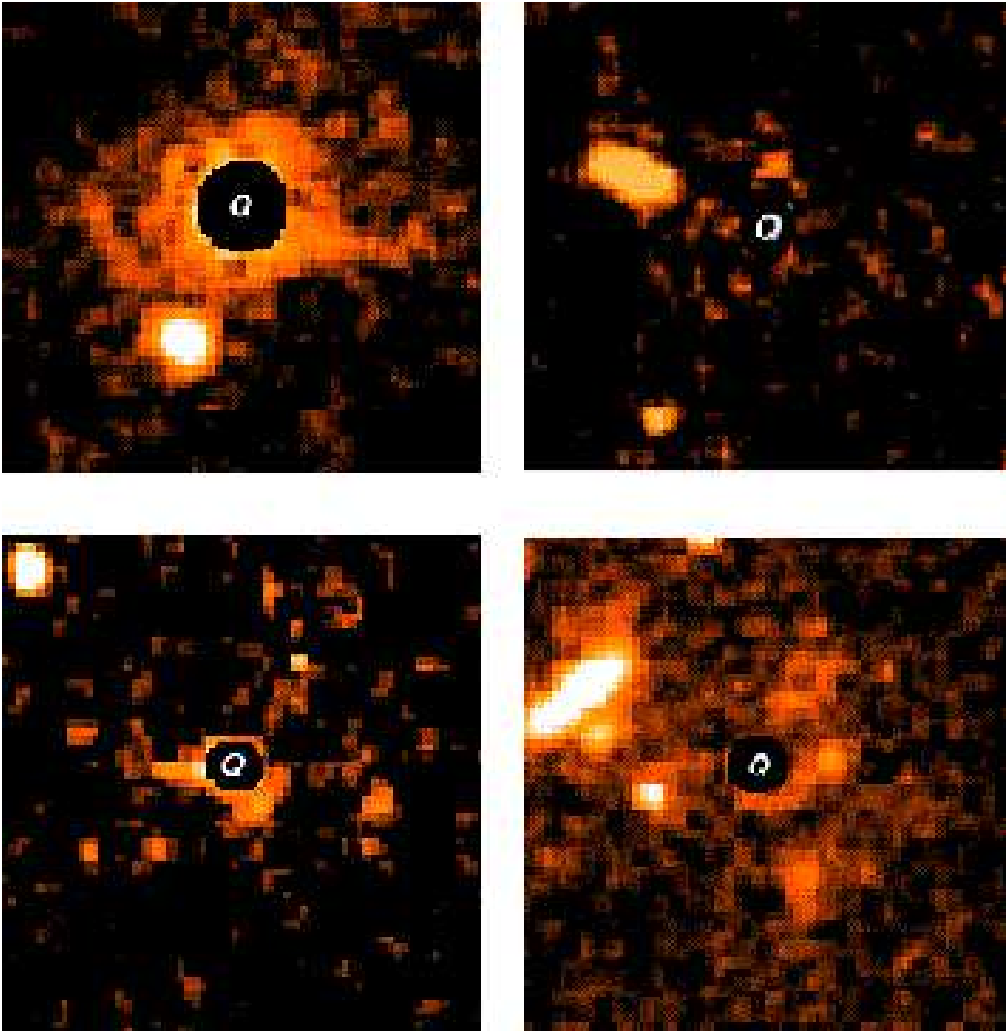}
\caption
{\small Four QSO fields with 5 DLA systems. The light at the position
of the QSOs (labeled Q) has been subtracted. North is up and east is
to the left.  {\bf Top left:} IRTF K band image ($\sim 18\arcsec
\times 18\arcsec$) of the Q0738+313 field.  The $z=0.091$ DLA galaxy
is identified to be the LSB galaxy nearly centered on the QSO. The
dwarf galaxy $\approx$ 6\arcsec\ to the southeast of the QSO has a
measured redshift of $z=0.22$. See Rao \& Turnshek (1998, ApJ, 500,
L115) and Turnshek et al. (2000, astro-ph/0010573)
for results on this field.  {\bf Top right:} IRTF K band
image ($\sim 20\arcsec \times 20\arcsec$) of the Q0827+243 field.  The
luminous galaxy $\approx$ 6\arcsec\ to the east-northeast has a
measured redshift of $z=0.52$ and is a probable spiral. Faint emission
due north of the QSO could be due to another galaxy although the
possibility of residuals from the QSO point spread function cannot be
ruled out.  {\bf Bottom left:} NTT J band image ($\sim 24\arcsec
\times 24\arcsec$) of the Q0952+179 field. The blobs to the east and
southwest of the QSO are LSB, but it is unclear if this is a
patchy/irregular structure or a single LSB galaxy.  {\bf Bottom
right:} NTT J band image ($\sim 27\arcsec \times 27\arcsec$) of the
Q1127$-$145 field. The DLA galaxy is associated with the
patchy/irregular LSB (north-to-south elongated) structure $\approx$
4\arcsec\ to the west; the brightest blob in this region has a
measured redshift of $z=0.31$.  }
\end{figure}

\end{document}